\documentclass[conference]{IEEEtran}
\usepackage{cite}
\usepackage{amsmath,amssymb,amsfonts}
\usepackage{algorithmic}
\usepackage{graphicx}
\usepackage{textcomp}
\usepackage{fancyhdr}
\usepackage{xcolor}
\usepackage{tikz}
\usepackage{amsmath}
\usepackage{hyperref}
\usepackage{multirow}
\usepackage{graphics}
\usepackage{enumitem}
\usepackage{color, colortbl}
\usepackage{lipsum}
\usepackage{tabularx,booktabs}
\usepackage[section]{placeins}
\usepackage{graphicx}
\usepackage{comment}
\usepackage[para]{footmisc}
\usepackage{rotating}
\usepackage{tablefootnote}
\usepackage{breakurl}
\usepackage[nobiblatex]{xurl}
\definecolor{lightgray}{rgb}{0.83, 0.83, 0.83}
\usepackage{caption}
\usepackage{subcaption}  
\usepackage{csquotes}

\ifCLASSINFOpdf
\else
\fi

\begin{document}

\title{What Are Brands Telling You About Smishing? A Cross-Industry Evaluation of Customer Guidance}

\author{
\IEEEauthorblockN{Dev Vikesh Doshi}
\IEEEauthorblockA{
California State University San Marcos\\
doshi002@csusm.edu}
\and
\IEEEauthorblockN{Mehjabeen Tasnim}
\IEEEauthorblockA{
California State University San Marcos\\
tasni001@csusm.edu}
\and
\IEEEauthorblockN{Fernando Landeros}
\IEEEauthorblockA{
California State University San Marcos\\
lande029@csusm.edu}
\and
\IEEEauthorblockN{Chinthagumpala Muni Venkatesh}
\IEEEauthorblockA{
California State University San Marcos\\
muniv005@csusm.edu}
\and
\IEEEauthorblockN{Daniel Timko}
\IEEEauthorblockA{
Emerging Threats Lab / Smishtank.com\\
danielttimko@gmail.com}
\and
\IEEEauthorblockN{Muhammad Lutfor Rahman}
\IEEEauthorblockA{
California State University San Marcos\\
mlrahman@csusm.edu}
}

\IEEEoverridecommandlockouts
\makeatletter\def\@IEEEpubidpullup{6.5\baselineskip}\makeatother
\IEEEpubid{\parbox{\columnwidth}{
		Symposium on Usable Security and Privacy (USEC) 2026 \\
		27 February 2026, San Diego, CA, USA \\
		ISBN 978-1-970672-07-7 \\
		https://dx.doi.org/10.14722/usec.2026.23067 \\
		www.ndss-symposium.org, https://www.usablesecurity.net/USEC/
}
\hspace{\columnsep}\makebox[\columnwidth]{}}

\maketitle

\begin{abstract}
Phishing attacks through text, also known as smishing, are a prevalent type of social engineering tactic in which attackers impersonate brands to deceive victims into providing personal information and/or money. While smishing awareness and cyber education are a key method by which organizations communicate this awareness, the guidance itself varies widely. In this paper, we investigate the state of practice of how 149 well-known brands across 25 categories educate their customers about smishing and what smishing prevention and reporting advice they provide. After conducting a comprehensive content analysis of the brands, we identified significant gaps in the smishing-related information provided: only 46\% of the 149 brands mentioned the definition of smishing, less than 1\% had a video tutorial on smishing, and only 50\% of brands provided instructions on how to report. Our study highlights variation in terminology, prevention advice, and reporting mechanisms across industries, with some brands recommending potentially ineffective strategies such as "ignoring suspicious messages." These findings establish a baseline for understanding the current state of industry smishing awareness advice and provide specific areas where standardization improvements are needed. From our evaluation, we provide recommendations for brands on how to offer streamlined education to their respective customers on smishing for better awareness and protection against increasing smishing attacks.

\end{abstract}


%
\IEEEpeerreviewmaketitle

\section{Introduction}

\hspace{0.4cm}
With the increased reliance on mobile devices over the past decade, there has been a dramatic rise in adversaries posing as reputable companies or representatives, and in customers falling victim to them~\cite{kroll2022report}. Yet public awareness of the threat has lagged behind; fewer than 35\% of the North American population is aware of what smishing is~\cite{safetydetectives_smishing}. In 2024, the FTC estimated that text message scams, including smishing, resulted in \$470 million in financial losses~\cite{keepnetlabs_smishing_stats}, a figure that is projected to continue growing~\cite{robokiller_mid_year_report}. Additionally, recent reports indicate that scammers are increasingly diversifying their targets, impersonating a wider range of organizations~\cite{apwg2025}.

\begin{figure}[htbp]
    \centering
    \includegraphics[width=0.7\linewidth]{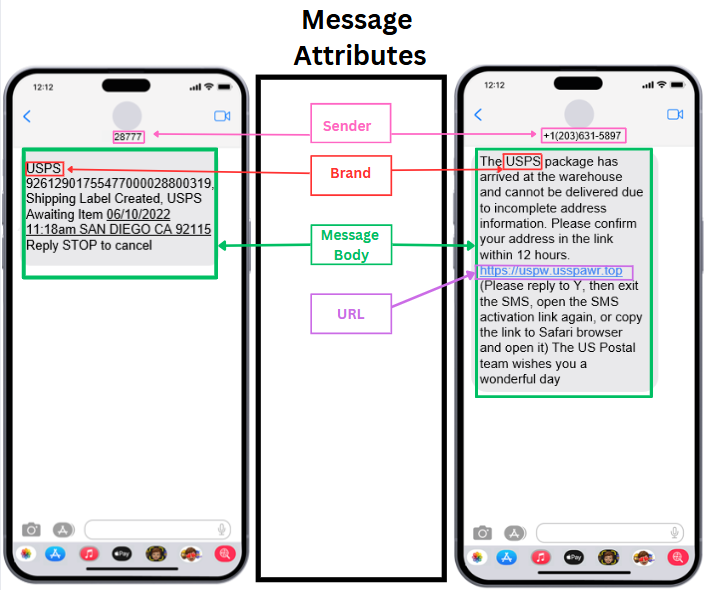}
    \caption{Screenshot of real (left) and fake (right) SMS of the same brand}
    \label{fig:SMSComparison}
\end{figure}

While phishing has been extensively studied in both academic and industry publications, smishing has received comparatively less attention, despite its growing prevalence~\cite{rahman2023users}. 
The unique threats that smishing poses can be seen in their delivery method, and in how users interact with messages on mobile devices~\cite{mahmud2024enhancing,yeboah2014phishing}. As shown in Figure~\ref{fig:SMSComparison}, taken from Smishtank~\cite{smishtank, timko2023commercial, timko2024smishing}, fake and real SMS from a brand can appear almost identical. With small changes to the URL (such as “uspw” instead of “usps”) or by using a phone number instead of a short code, users who lack technical knowledge or awareness of smishing are often tricked into believing the messages come from legitimate brands. 

Given these clear risks to end users, many companies have publicly share anti-smishing awareness advice~\cite{amazon, walmart, morgan, verizon, viasat, toyota}. These efforts aim to mitigate the risk of falling prey to attackers. While most advice is geared towards employees rather than customers in order to prevent data breaches~\cite{cisaemployee,ncscemployee}, brands also release information for their end users, which can help avoid reputational damage and indirect costs such as customer support and loss of business~\cite{BOSE201467}. 
This customer-facing advice takes many forms, including videos, text, and multimedia instruction on how to spot an attack, as well as what to do when you identify an attack. With each company taking their own unique approach, the need arises to investigate the nature of the anti-smishing information given by brands and the tools they offer to end users to protect themselves from fraud~\cite{mossano2020analysis}. This advice forms a key aspect of cybersecurity education. By examining how this advice is communicated and where it may fall short, we can identify areas where brands can better protect end users and their trust in the brand.

Cybersecurity education is widely recognized as key to preventing cybersecurity attacks, and brands can play a crucial role in this educational process~\cite{alghamdi2017can,reinheimer2020investigation,kweon2021utility}.
Given their direct customer relationships and the trust they command, brands have a unique opportunity, and responsibility, to offer accurate anti-smishing guidance, particularly in light of the potential consequences of failing to address such attacks~\cite{jampen2020don}. 
For example, according to a 2019 U.S. consumer survey commissioned by the brand-protection firm Incopro, 52\% of respondents reported losing trust in a brand after unintentionally purchasing a counterfeit good online~\cite{Incorpo}. Moreover, a brand-trust survey by Mimecast found that 57\% of respondents stated they would stop spending money with their favorite brand if they fell victim to a phishing attack impersonating that brand~\cite{Mimecast}.

As primary targets of smishing attacks, brands are well-positioned to educate customers using their platform to provide tailored, trustworthy guidance. 
To understand the current state of industry practice for smishing awareness and advice, we conducted the first systematic content analysis of anti-smishing advice from major corporations, examining common themes, use of supplementary information, and post-incident guidance approaches. This foundational study aims to document what information and guidance brands currently provide, establishing a baseline for understanding industry practices. Our research questions are:
\begin{enumerate}
    \item \textbf{RQ1: }  How do brands define smishing and the methods to identify smishing attacks?
    
    \item \textbf{RQ2:} What percentage of brands provide additional information such as external links, videos, or examples?
    
    \item \textbf{RQ3:} What instructions do brands provide to prevent smishing, report a smish attempt, and what steps to take after becoming a victim of smishing?

\end{enumerate}

To answer these research questions, we identified 149 popular brands from Smishtank~\cite{smishtank, timko2023commercial, timko2024smishing} and Phishtank~\cite{phishtank} and collected data from the websites of those 149 brands. We then performed a qualitative analysis of the advice offered by brands to educate their end users.
We found that less than 50\% of the brands mentioned smishing on their websites. Only 35\% of the brands provided smishing examples, and less than 1\% of the brands had video tutorials on smishing for their users. Brands defined smishing as an act of deception involving malicious links. To prevent their users from falling victim to a smishing attack, brands advised their users to trust their instincts and block the smishing contact.

Based on these findings, we recommended that brands standardize smishing definitions for better understanding. We also recommend mentioning simplified reporting methods and including government and carrier resources so users do not have to navigate each brand's specific reporting process. Another recommendation is for brands to invest in modern multimedia to educate their users about smishing attacks.

\section{Related Works}
\label{sec:relatedworks}

\textbf{Smishing Related Studies}
While smishing research remains limited, recent studies have shed light on user susceptibility to these SMS phishing attacks. Timko et al. found that US users, on average, achieved 65.60\% accuracy in identifying fake SMSes compared to 44.6\% for real messages~\cite{timko2025understanding}. Worryingly, response rates to smishing messages exceeded 15\%, especially when senders posed as known brands or organizations. Given that the number of people falling for cyber attacks is increasing~\cite{smishincrease}, new methods are constantly being implemented with the goal of reducing that number. For example, Sheng et al.~\cite{sheng2007anti} experimented with an online game that provided anti-phishing training to users. Toll scams, which often begin with SMS phishing, have also emerged as a threat. Recent work~\cite{munny2025infrastructure} analyzed thousands of scam domains, revealing registrar abuse, obscure TLD use, and coordinated registration bursts.

\textbf{User Behavior} Beyond detection accuracy, user behavior plays a crucial role in smishing success. Blancaflor et al. identified a “curiosity trap," where users tend to open suspicious messages to verify legitimacy, inadvertently exposing themselves to malicious content~\cite{blancaflor2021phishing}. This behavior contributes to the 4.17\% success rate of smishing campaigns. The blurring line between real and fake messages further worsens the issue, as users ignore 50\% of genuine messages due to perceived smishing~\cite{jamesTeenagersAbility}. According to a study conducted by Neupane et al., users subconsciously process and identify aspects of real and fake links differently~\cite{neupane2015multi}. Similarly, Tabassum et al. explored factors that drive smishing susceptibility, examining how and why users judge a SMS message to be real or fake. These studies highlight the importance of message design and user education in differentiating between legitimate and fraudulent communications. Furthermore, a past study found that while many users do not recognize the term smishing, they are aware of the risk that comes with being a victim of an SMS attack~\cite{Edwards2023vector}. Despite this, Downs et al. explained in their study that users have a hard time linking this awareness to their susceptibility to an attack~\cite{downs2006decision}. These findings underscore the critical importance of clear, accessible organizational communication about smishing threats.

Research conducted by Karakasiliotis et al. concluded that cyber attacks have both psychological and technological dimensions~\cite{Karakasiliotis2006aware}. Li et al. identified trust building as a crucial step for a phishing attack to succeed~\cite{Li2023pattern}. Alarmingly, Camp et al. found that many users trust their computer systems to the extent that they underestimate risks, which could explain why many users click on visually deceptive links~\cite{Camp2009mental}. A study by Cho et al. explored how various personality traits impact a user's willingness to trust someone~\cite{Cho2016trait}. They found that users with high levels of anxiety, i.e., neuroticism, were less likely to trust what they were sent and, in turn, were more likely to consider the associated risks. Attackers employ various techniques to deceive users, making it difficult to distinguish genuine messages from fraudulent ones. Drake et al. described a technological method for this by masking a URL address to appear as if it is sourced from a reputable website~\cite{Drake2004anatomy}. Another deception technique is emotional manipulation. Goel et al.~\cite{goel2017got} demonstrated that emotional manipulation can influence users' decision-making by eliciting feelings of fear and anticipation, proving emotions to be an effective way of getting someone to click on a harmful link. This extensive body of research highlights significant user behavioral vulnerabilities to smishing attacks. While previous studies have thoroughly demonstrated what makes users vulnerable to smishing attacks, there is limited understanding of what information organizations currently provide to customers about these threats. Our study systematically documents current organizational practices by presenting an evaluation of smishing education content and guidance that brands currently offer to help users recognize and avoid the behavioral manipulation tactics employed in smishing attacks.

\textbf{How Brands Provide User Cyber Education}
With the escalation of smishing in recent years, the number of scammers impersonating reputable and trustworthy brands to deceive users is also increasing~\cite{proofpoint2023statephish}. These kinds of attacks, where scammers pretend to be well-known brands, are becoming very common, and these scammers are becoming more advanced day by day. A report by the Federal Trade Commission noted an 85\% increase in brand impersonation between October 2020 and September 2021, which resulted in losses of USD 2 billion~\cite{ftc_impersonation_fraud}. Additionally, a report highlighted that eight in ten organizations (i.e., 80\%) experienced at least one successful email-based phishing attack in 2022, which further emphasizes the critical need for enhanced cybersecurity education and awareness~\cite{proofpoint2023statephish}.

Cybersecurity education has become an essential tool in response to these threats, reducing user susceptibility to smishing. Education directed at improving user detection takes a wide variety of forms, and targeted training programs have been developed to help vulnerable populations. A study by Lastdrager et al. found that targeted training towards children and teens significantly improved their detection accuracy through training~\cite{lastdrager2017effective}. In a different approach, Wang et al.~\cite{wang2025can} found that cybersecurity education which employed evidence-based AI explanations for message detection was able to improve user detection of smishing across all age groups, with the greatest improvement in older adults. Effective cybersecurity education has the potential to equip users with the necessary knowledge and skills to detect and avoid cyber threats. Many research studies emphasize the significance of adopting effective, learning-based approaches to cybersecurity training. Bada et al. mention that educational efforts should go beyond merely providing information; they should be targeted, actionable, and practical, offering clear steps that users can implement immediately~\cite{bada2019cybersecurityawarenesscampaigns}. Additionally, Bhaskar et al. state that showing the aftermath of real breaches and highlighting how vulnerable organizations are to cyber threats is the best possible way to spread cybersecurity awareness~\cite{bhaskar2022better}.

In addition to traditional methods, brands have started adopting more user-centric cybersecurity education techniques, which include interactive models and game-based learning platforms. A study by Williams et al. demonstrated that a game-based approach in cybersecurity education effectively engages and inspires non-cyber students~\cite{williams2024leveraging}. Similarly, research by Khan et al. highlighted the benefits of game-based learning platforms in enhancing cybersecurity education~\cite{khan2022game}. Organizations like Google have implemented free phishing quizzes to assess and improve user awareness~\cite{googlequiz}.

Despite all the efforts, there is a significant gap in understanding how different brands educate their users about cybersecurity, especially smishing. Our research study aims to fill this gap by conducting an extensive content analysis study of 149 brands. We examine whether brands spread awareness about smishing, what awareness they provide, and how they want their users to approach smishing attacks based on their unique brand strategies. By systematically cataloging current approaches, we seek to establish a baseline understanding of industry practices and identify the range of strategies currently employed. We also aim to document common patterns and approaches found across different organizations' smishing education materials. Through this analysis, we aim to map the current landscape of organizational anti-smishing communications and identify gaps where industry practice could be strengthened or standardized.

\section{Methodology}

This study analyzes anti-smishing cybersecurity advice provided by 149 U.S. brands that are highly represented in public phishing and smishing datasets. To assess both the content and comprehensiveness of corporate guidance, we employed a mixed-methods approach combining systematic content collection, qualitative coding analysis to identify thematic patterns in advice, and quantitative frequency analysis to measure the prevalence of educational elements that can help prepare users for smishing attacks.

We collected data across eight categories addressing the full lifecycle of user engagement. Five categories captured text content for qualitative analysis of the advice and guidance regarding smishing, while three used binary coding for quantitative analysis to measure the presence of smishing-related resources. For each brand, we systematically searched official websites for smishing-related guidance. We performed qualitative analysis using open coding techniques in ATLAS.ti, with AI-assisted initial coding followed by rigorous manual validation and iterative refinement.

The methodology includes brand selection from Smishtank and Phishtank databases based on frequency of impersonation reports, systematic data collection from official brand websites, and codebook development through hybrid AI-assisted and manual coding processes.

\begin{table*}[h!]
\centering
\begin{tabular}{|m{4cm}|m{12cm}|}
\hline
\textbf{Categories} & \textbf{Description} \\ \hline
Definition of smishing & The company’s definition of smishing, including definitions of phishing via text message or SMS. \\ \hline
How to identify smishing & Details commonly found in smishing messages. \\ \hline
Is there a video tutorial provided? & If a video tutorial on smishing is provided. \\ \hline
Are there examples of smishing provided? & If explicit examples of what a smishing message might look like are provided. \\ \hline
Are there any external links provided? & If external links are mentioned for further help or guidelines. \\ \hline
Steps on how to report fraud & Instructions on how to report a suspicious text message. \\ \hline
Steps to take after becoming a victim & Instructions for smishing victims to mitigate damages. \\ \hline
Smishing prevention advice & Advice on preventing customers from becoming victims of smishing. \\ \hline

\end{tabular}
\caption{Description of categories}
\label{Tab:Themes}
\end{table*}

 \begin{table}[h]
\centering
\small 
\begin{tabular}{|l|c|}
\hline
\textbf{Category} & \textbf{Count} \\ \hline
Technology \& Software & 25 \\ \hline
Retail \& E-commerce & 23 \\ \hline
Finance \& Banking & 23 \\ \hline
Telecommunications & 7 \\ \hline
Social Media & 6 \\ \hline
Automotive \& Ground Travel & 6 \\ \hline
Cybersecurity \& VPN & 6 \\ \hline
Food \& Beverages & 5 \\ \hline
Gaming & 5 \\ \hline
Payment Services & 5 \\ \hline
Logistics \& Delivery & 4 \\ \hline
Health \& Pharmaceuticals & 4 \\ \hline
Insurance & 4 \\ \hline
Housing \& Utilities & 4 \\ \hline
Entertainment/Streaming & 4 \\ \hline
Dating Platform & 3 \\ \hline
Air Travel & 3 \\ \hline
Government & 3 \\ \hline
News \& Information & 2 \\ \hline
Finance \& Investments & 2 \\ \hline
Cryptocurrency & 2 \\ \hline
Education & 1 \\ \hline
Non-Profit \& Advocacy & 1 \\ \hline
Food Delivery & 1 \\ \hline
\end{tabular}
\caption{Distribution of the brands}
\label{Tab:BrandsDistribution}
\end{table}

\subsection{Brand selection} 
 We selected brands from phishtank.com~\cite{phishtank} and smishtank.com~\cite{smishtank, timko2023commercial, timko2024smishing}. Both of these websites are known for reporting phishing and smishing attempts, respectively. We identified the most reported brands on these two websites and only selected the brands that are popular in the United States region. Choosing frequently reported brands allows us to cover the brands that are commonly impersonated. Their popularity was determined by the number of phishing reports linked to each brand. 

We selected a total of 149 distinct brands to have a diverse and healthy range of data. 
The brands selected were from a wide range of categories, including but not limited to Technology \& Software, Retail \& E-commerce, Social Media, and Food and Delivery (mentioned in detail in \textit{Table \ref{Tab:BrandsDistribution}}). Choosing from a diverse range ensured that we analyzed brands across different sectors. The full list of all brands chosen for each category can be found in the appendix.

 \subsection{Data collection}
For the data collection process, we included eight distinct categories designed to answer our three research questions. The categories included:
\begin{enumerate}
    \itemsep0em 
  \item Definition of smishing (\textbf{RQ1})
    \item How to identify smishing (\textbf{RQ1})
    \item Are there video tutorials provided? (\textbf{RQ2})
    \item Are there examples of smishing provided? (\textbf{RQ2})
    \item Are there any external links provided? (\textbf{RQ2})
    \item Steps on how to report fraud (\textbf{RQ3})
    \item Steps to take after becoming a victim (\textbf{RQ3})
    \item Smishing prevention advice (\textbf{RQ3})
    
\end{enumerate}

The choice of coding topics reflects major challenges across the full lifecycle of user engagement with smishing threats. That being, what is smishing, how you identify it, how to mitigate it, how to report it, and what to do if you have fallen victim. Researchers have made strong cases for the importance of each of these topics~\cite{lastdrager2014achieving,naqvi2023mitigation,10.1145/3577923.3583640}. Additionally, these choices closely align with categories of focus by CISA~\cite{cisa_2021}.

These categories (described more in \textit{Table \ref{Tab:Themes}}) ensured that we collected enough information to correctly analyze and answer our research questions in a structured manner. Each of the categories covered a distinct aspect of this study. For example, “Steps to take after becoming a victim" and “Steps on how to report fraud instructions" were designed to answer \textbf{RQ3} (What instructions do brands provide to prevent smishing, report a smish attempt, and what steps to take after becoming a victim of smishing?).

We analyzed the websites of each of the 149 brands thoroughly. This included visiting the website of each brand, and searching for the smishing and smishing-related terms to find any help or guidelines provided. The search terms included smishing, scam or spam texts, SMS fraud, and SMS phishing. We also used Google site-specific searches with these phrases (e.g., 'site:brandX.com smishing') to identify smishing-related pages within each brand's domain. If the brand's website did not have a search feature,  we manually reviewed the FAQ, Help, and About pages of the brand to look for smishing guidelines. If we did find any information or guidelines related to smishing, we copied it into our dataset. 

We maintained an Excel sheet to track and store our data. We had one column for each of the above-mentioned eight categories and additional columns for the brand name, link to the website where the information was provided, and category of the company. Each set of information found on a brand's website was mapped to its respective column. For the categories, “examples of smishing," “video tutorials provided," and “external links included," we recorded binary values ('Yes' or 'None') to enable quantitative analysis of prevalence across brands. For the remaining five categories, we collected the full text of the category content from brand websites for subsequent qualitative coding analysis.

\subsection{Codebook Development}

After data collection, we conducted a qualitative analysis using open coding techniques~\cite{flick2013sage}. ATLAS.ti was used to support dataset organization and the generation of preliminary candidate codes~\cite{Atlas}, leveraging its integration of fine-tuned OpenAI models~\cite{atlasti_why}. Two researchers reviewed and manually refined the candidate codes into their final form, discussing emerging themes and collaboratively developing the codebook. Discrepancies in the codes were resolved through collaborative discussion supporting alignment in interpretations and reducing potential bias.  This hybrid approach enhance the rigor and reliability of our analysis.

While software-assisted coding differs from traditional manual open coding methods, Computer-Assisted Qualitative Data Analysis Software (CAQDAS) tools are widely used in qualitative research, particularly when combined with systematic human validation~\cite{abumalloh2024exploring,mohamad2022inductive}. The full codebook is provided in the appendix.

\section{Results}

\subsection{Answering RQ1}
\subsubsection{Smishing Definitions}
The study identified several themes regarding smishing techniques and their characteristics. Deceptive messaging emerged as a prominent strategy, with brands frequently highlighting messages that impersonated brands or persons to appear legitimate. As one brand mentioned,

\begin{displayquote}
    \textit{“One way phishers try to gain this information is by posing as a reliable company or person in emails, texts or direct messages. They will send some form of communication asking for verification, information updates, financial updates or even just try to get you to navigate to a provided link. (Yahoo)"}
\end{displayquote}

This aligns with the theme of requesting personal information and embedding malicious links, with such tactics designed to manipulate users' trust and urgency.

Another recurring theme was the creation of urgency, with messages instructing users to “act immediately" or risk consequences. For example, a brand shared,

\begin{displayquote}
    “\textit{Creating a sense of urgency to pressure the recipient into taking immediate action. (ExxonMobil)"}
\end{displayquote}

Smishing was also recognized as a type of social engineering attack, often involving short URLs or trustworthy entities to obfuscate the malicious intent. Brands described these tactics as a manipulative approach to electronic fraud activity, emphasizing the psychological tricks used to extract information or provoke action. One brand summarized this succinctly, stating,
\begin{displayquote}
    “\textit{Smishing is a form of social engineering attack where attackers use SMS (Short Message Service) or text messages to trick individuals into divulging sensitive information, clicking on malicious links, or taking other malicious actions. (Berkshire Hathaway)"}
\end{displayquote}

Some brands also refer to smishing as SMS phishing, as phishing is a more widely known term, as stated by one of the brands,
\begin{displayquote}
    “\textit{Smishing, or SMS phishing, is a phishing technique. (NordVPN)"}
\end{displayquote}
A brand also defined smishing as a technique to install malware through malicious links and then further extract a person's personal information. For example,
\begin{displayquote}
    “\textit{often similar to texts you might receive from legitimate businesses. If you click the fake link, you might be asked for personal information or get malware installed on your phone that can extract information automatically. (Navy Federal Credit Union)}"
\end{displayquote}

Another theme that we observed was defining smishing as an SMS that used fear as a strategy to trap a victim. This tactic would make the participants more vulnerable to falling into the smish trap. As mentioned by a brand, 
\begin{displayquote}
    “\textit{Urgent or Threatening Language: Smishing messages often create a sense of urgency or contain threats to prompt immediate action. (IBM)"}
\end{displayquote}
Adding to this, a brand also defined smishing as a text message that creates a sense of urgency, preventing the victim from taking the time to make an informed decision. For example, a brand wrote,
\begin{displayquote}
    “\textit{Smishing messages often create a sense of urgency or include threats to prompt immediate action. (Amazon)"}
\end{displayquote}

Out of the 149 brands that we analyzed, we found that only 46\% of brands defined smishing. Each of the 69 brands that defined smishing had similar definitions as described in the above themes. However, it is important to note that more than 50\% of brands did not even mention what smishing is. This is a significant concern, as there have been instances where smishing attempts have been made by attackers pretending to be the brand. When users visit the websites of brands that do not mention smishing, they have no idea about smishing.
\subsubsection{How to Identify Smishing}
The study highlighted some critical themes on how scammers operate during a smishing attack and the signs of identifying them. The most recurring theme was asking for personal information. Most brands warned users that requests for sensitive data such as credit card numbers, bank account details, PINs, CVV codes, one-time passwords, or even full names can be a clear indication of smishing.

Another significant theme was the sense of urgency in the smishing text. Scammers use these tactics to pressure victims into acting quickly, giving them less time to contemplate their actions. These urgent requests are usually to log in, send money, or confirm account details. As one brand explained,
\begin{displayquote}
    \textit{“Urgent or Alarming Content: Smishing messages often convey urgency or alarm to prompt immediate action (Coco Cola)."}
\end{displayquote}
Unexpected links or attachments from senders were also seen as a common theme in this study. If the text contains unusually lengthy links or unexpected attachments, that is a big red flag. As one brand noted,
\begin{displayquote}
    \textit{“Phishing emails may include attachments claiming to be a 1099 tax document or other important files (Robinhood)."}
\end{displayquote}
These links in the SMS sometimes mismatch with the official service provider's email. Even though the links are usually lengthy, it does not mean short URLs are secure. Often, scammers shorten URLs to mislead people. To combat this, brands recommend hovering over any link to preview the URL without clicking, as highlighted in the advice:
\begin{displayquote}
    \textit{“If the message includes a link, hover over it (without clicking) to preview the URL. Verify that the URL is legitimate and related to the claimed sender (Walmart)."}
\end{displayquote}
Misspelled websites or email addresses in the SMS are another big red flag. Communications from any legitimate company are well-written and error-free. Misspellings, poor grammar, inconsistent fonts, or odd markings are clear indicators of smishing attempts. As one brand highlighted,
\begin{displayquote}
    \textit{“Lack of proper grammar and punctuation, many of these scams originate from foreign countries(Consumer Cellular)"}
\end{displayquote}
Trusting the user's instincts is also a major theme found in the study. If a message seems even slightly suspicious, the user should stop engaging and think before taking any further steps. As one brand emphasized, 
\begin{displayquote}
    \textit{“If something feels off or  too good to be true, it might be a smishing attempt (Walmart)."}
\end{displayquote}
Scammers often lure victims with claims of winning large sums of money, sponsored lotteries, fake prize notifications, gift cards, or promotions. These tempting offers are made to attract people, as mentioned,
\begin{displayquote}
    \textit{“Scammers may also contact you with attractive offers for free stock or other enticing deals to lure you in (Robinhood)."}
\end{displayquote}
Requests to download apps, install software, or be directed to third-party websites are additional indicators of smishing. Scammers attempt to gain access by provoking victims to install malware on their systems. Therefore, any request to download a third-party app or involve a different party for any kind of activity is highly suspicious and a clear sign of smishing, as a brand mentioned,
\begin{displayquote}
    \textit{“Hulu will not direct you to an unfamiliar third party for support. We will never direct you to install third-party software in order to troubleshoot (Hulu)."}
\end{displayquote}

Requesting funds for various activities is another common scamming tactic. Scammers ask for payments, claiming they will apply for a job, pay insurance or delivery fees, or provide administrative fees for grant funds. Usually, they include a third party to fulfill the payment, making it an obvious scam.

Brands have mentioned that scammers try to pretend to be a brand that most people trust. They often impersonate well-known service providers like Instagram, claiming to be from their security team to extract account information. They manipulate the links to make them look like the authentic service provider's email address or phone number, so victims get confused. For example,
\begin{displayquote}
    \textit{“The sender’s email or phone doesn’t match the name of the company that it claims to be from (Apple)."}
\end{displayquote}

Scammers also use emotional manipulation techniques. As one brand mentioned,
\begin{displayquote}
    \textit{“Emotional appeals—attempts to lean on a personal connection, either new or preexisting, can be a way to get you to stop thinking rationally and leave you vulnerable (Discover)."}
\end{displayquote}
They sometimes behave overly friendly but make persistent requests. On the other hand, they may impose threats to instill fear in victims, leading them to take immediate but wrong actions and fall for the scam. One brand noted,
\begin{displayquote}
    \textit{“Alarming threats of drastic consequences involving financial and legal issues if there is not an immediate response (Consumer Cellular)."}
\end{displayquote}

Out of the 149 brands analyzed, 57\% (85 brands) did not provide any instructions on how to identify smishing attempts. Only 43\% (64 brands) offered guidance, leaving a significant portion of users without the necessary resources to recognize and avoid these scams. This gap in support is significant given that we these are popular brands that have already been targeted. This absence of smishing-specific educational resources not only makes users more vulnerable, but also highlights the importance of providing a standardized guidelines to avoid confusion between end-users. Without a consistent standard approach to educational resources, users are left to distinguish their legitimate messaging from the spoofed version on their own, raising the risk of exploitation. 

\textbf{RQ1 Summary:} Different brands have their own ways of defining smishing. The spectrum of definitions ranges from smishing being an act of deception to impersonating a brand or person to SMS involving short links. For providing guidelines to identify smishing attacks, some brands mentioned noticing technical details like checking the link address, sender's details, and if the SMS contains any attachments, whereas other brands mentioned trusting personal capabilities such as following emotional appeal.

\subsection{Answering RQ2}
\subsubsection{Video Tutorial}
We also analyzed how many brands provided a video tutorial on smishing. Videos can depict a better scenario of what a smish message looks like, what a malicious link looks like, and what red flags to look for while examining an SMS. Moreover, these video tutorials can also walk through the process of identifying a smish, correctly reporting and blocking the sender, and the steps to take after becoming a victim. Videos offer a multimodal learning experience by integrating text, audio, and visual elements, which can enhance engagement and is preferred by some users when compared to text-only content~\cite{abawajy2014user,volkamer2018developing}.

We found that only 1 out of the 149 brands we explored provided a video tutorial on smishing. This indicates a lack of interactive videos and examples mentioned by the brands on their websites. This also highlights and encourages scammers to use different tools and leverage the lack of awareness to scam more potential victims. Exploring modern multimedia tools and providing cybersecurity awareness in an interactive way could strongly portray the brand's commitment to the safety of its customers, as even suggested by Bada et al. \cite{bada2019cybersecurityawarenesscampaigns}.

\subsubsection{Smishing examples}

We explored how many brands provide a smishing example on their website. Smishing examples can explain a smish in a digital way compared to text content. These examples would help customers understand what a smish looks like and depict scenarios used by attackers. Seeing examples would also help potential victims understand the specific scenarios of urgency or fear a smish could create. For example, a bank could have smish examples of fake scenarios like payment failures or fraudulent transactions.

Only 35\% of the 149 brands we explored mentioned a smishing example or screenshot. This indicates a significant gap in educating customers about the specific smishing scenarios of the respective brands. It also highlights the brands' failure to utilize modern multimedia options to better explain a cyber attack rather than just writing text. Providing examples and screenshots can greatly aid user understanding of smishing scenarios and potential attacks specific to each brand.

\subsubsection{External Links}
Out of the 149 brands we explored, 47 brands provided external links as additional resources for their users. These external links vary from personal websites of the respective brands to links from sources such as law enforcement agencies. However, less than 50\% of the brands provided an external link reference, the brands provided some valuable extra information on smishing for the user.

These links provided users with further guidance and support in case they did not have enough knowledge about the aftermath of the smishing. Some brands (for example, Instagram) direct the users to their official support page which has tips about keeping your account safe. Some brands direct to blogs or posts posted by cybersecurity organizations on how to protect user privacy online. By providing these links, brands not only educate and prepare their user to be safe from scammers but also give them useful resources to equip them with proper tools in case the users need further help to report the smishing.

\textbf{RQ2 summary:} We observed less than 50\% of the brands provided additional and interactive information such as external links, video tutorials, and smishing examples. However, the brands that did not provide any additional information, listed valuable insights and helpful resources for their user. For example, a few brands provided screenshots of previously reported smishing scams for the user to have a better understanding. Some brands even provided links to government helplines for best reporting practices and official guidelines to follow concerning smishing. Overall, a small proportion of brands provided additional information but those who did provide had mentioned important and true resources. 

\subsection{Answering RQ3}
\subsubsection{Reporting Instructions}
The study identified several key themes regarding instructions for reporting fraud. Reporting to the respective brand was the most emphasized approach, with brands frequently noting how this step ensures that responsible brands are informed promptly. This also ensures, that the smish attempt is correctly reported and the required steps are taken to avoid a future scam. Furthermore, this suggests that respective brands have different ways to block and report the sender as per the type of brand, so it is best to refer to and apply the methods defined by the respective brand. As one brand mentioned,
\begin{displayquote}
  “\textit{If you see something that you believe is a scam, avoid responding or interacting with it and report it to Instagram immediately. (Instagram)"}  
\end{displayquote}

Another significant theme was the role of government authorities and specialized agencies. Brands often mentioned forwarding fraudulent messages to entities like the FTC, FBI, or the cybercrime division, with one brand sharing, 
\begin{displayquote}
    “\textit{Consider filing a report with the Federal Trade Commission and/or state attorney general’s consumer protection office, or the FBI. (Costco)"}
\end{displayquote}
Additionally, using designated numbers like 7726 was highlighted as a quick and effective way to alert service providers about spam or phishing attempts.

The inclusion of specific details in reports also emerged as a key theme. Brands described how providing timestamps, screenshots, and scam details added credibility and clarity to their reports. For instance, one brand mentioned, 
\begin{displayquote}
    “\textit{Text message: Screenshot the message and include the number that contacted you Phone call: Include the phone number from the call and share as much detail as possible (Robinhood)"}
\end{displayquote}
 This approach aligns with the importance of utilizing online reporting platforms and official channels to streamline the reporting process.

Finally, brands noted preventive measures such as marking fraudulent messages as spam and deleting them. One brand summarized this sentiment, stating, 
\begin{displayquote}
    “\textit{Report spam in Google Messages When you report a conversation as spam, you also block the sender and move the message to your "Spam \& blocked" folder. (Google)"}
\end{displayquote}
 Some brands even mentioned educating others about the scam and spreading awareness so people are notified and alerted. As one brand stated 
 \begin{displayquote}
     “\textit{Alert Your Contacts: If the fraudulent instructions involve requests to your contacts or associates, inform them about the potential scam. This helps prevent the spread of the scam to others. (Walmart)"} 
 \end{displayquote}

Out of the 149 brands that we analyzed, we found that only 50\% of the brands provided steps on how to report a smishing scam. The other 50\% are popular brands that have no instructions defined on how to report a smish attempt. This also supports the fact that lack of reporting which further creates a lack of awareness is one of the common reasons for smishing scammers to survive and scam more and more victims~\cite{securityeducation}. 

\subsubsection{Steps After Becoming a Victim}

The study highlighted some core themes on what to do after becoming a victim of a smishing attempt. Almost all the brands ask the user to call or inform them immediately if the users think they have been scammed. If any bank information was shared with the scammer or there was a financial loss of any kind, brands highly suggest contacting the bank immediately and getting help from them. Some brands also ask the victims to report the scam and forward the message to the mobile carriers. As one brand stated,
\begin{displayquote}
    \textit{“If possible, forward the smishing message to your mobile service provider, as they may use it to improve their security. (IBM)"}
\end{displayquote}

Brands also encourage the victims to report the scams to appropriate law enforcement agencies. Filing a complaint to the Federal Trade Commission (FTC) is the most common law enforcement agency that was brought up by those brands. Some other agencies that were mentioned are the Internet Crime Complaint Center, the state Attorney General's office, local cyber-crime unit. local or state police agencies. In the case of any identity theft, one brand suggested filing an affidavit at www.identitytheft.com.

Another significant theme was to change their password to a stronger one and incorporate uniqueness for any accounts that were compromised and then log out from all devices. Brands also recommend to not use the compromised device if possible just to be cautious about any malware. One brand mentioned,
\begin{displayquote}
    \textit{“Stop using the device. In fact, don't even use the affected device (laptop, phone) to change your passwords. It may be infected with malware that will collect your new passwords, leaving you no better off than before (Synchrony)."}
\end{displayquote}

Regular monitoring of the accounts is also one of the crucial themes found in the study. The accounts that have been compromised before to detect any suspicious activity. Monitoring financial accounts was emphasized for any unrecognized transactions. 

 After becoming a victim, scanning the device with an anti-virus is necessary to be protected from any possible malware. If any anti-virus is not already installed, brands recommend installing a reliable anti-virus and scanning the infected device with it as soon as possible.

 At any point, if there is any other text that seems to be suspicious, do not engage.  Brands also say not to click any links that do not look reliable. When interaction has already been established, cease any communication immediately because replying can trigger future smishing attempts as the scammers know the number is active. In that situation, brands ask victims to stop using the device. For example,
 \begin{displayquote}
    \textit{“Then, disconnect the potentially infected device completely from the internet and cell service, which will help isolate any issues. (Synchrony)"}
\end{displayquote}

For an extra level of security, several brands emphasize enabling two-factor verification whenever possible for accounts. As one brand mentioned,
 \begin{displayquote}
    \textit{“This adds an extra layer of security by requiring a second form of verification in addition to your password. (Tech Mahindra)"}
\end{displayquote}

Finally, brands noted some general measures such as information from family, friends, and colleagues so scammers cannot harm them. Being calm and collected is important so the situation does not get worse. Blocking those smishing phone numbers is crucial. It was also recommended to learn about cybersecurity and do some research online to be aware of safety in the future.

Out of the 149 brands analyzed, 65.77\% (98 brands) did not provide any clear steps for users to follow after becoming a victim of smishing scams. Only 34.23\% (51 brands) offered guidance, leaving a majority of users without critical support during such incidents. This underscores a significant gap in victim assistance and highlights the urgent need for brands to prioritize victim-focused measures to address this vulnerability.

\subsubsection{Smishing Prevention Advice}
The study identified a range of strategies for preventing smishing attacks, with an emphasis on user vigilance and proactive security measures. A prominent theme was avoiding interactions with suspicious content, particularly refraining from clicking links or opening attachments, which was frequently cited by participants. As one brand explained, 
\begin{displayquote}
    “\textit{Don't click a link if you're not sure about it; go directly to the company website instead. (Netflix)"}
\end{displayquote}
Similarly, exercising caution when engaging with SMS and verifying the legitimacy of senders emerged as critical preventive measures, highlighting the need for skepticism.

Another significant theme was the protection of personal information, with brands emphasizing the importance of not sharing sensitive data such as account credentials or financial details. For instance, one brand noted, 
\begin{displayquote}
    “\textit{Never provide personal or financial information unless you are certain of the identity of the person or business that is contacting you. (Walgreens)"}
\end{displayquote}
Brands also stressed the necessity of verifying sender authenticity multiple times before transferring funds, underscoring a measured and cautious approach to potential threats.

Technical safeguards were another recurring theme. Installing the latest antivirus software, enabling two-factor authentication, and maintaining strong passwords were commonly recommended actions. Brands highlighted the value of such precautions, with one noting, 
\begin{displayquote}
    “\textit{Make sure your devices and security software are up to date. Consider installing anti-malware software for added security. (Capital One)"}
\end{displayquote}
Additionally, double-checking website URLs and relying only on official apps or platforms for transactions were identified as critical strategies to avoid SMS phishing traps.

Brands also highlighted the importance of proactive measures, such as reporting suspicious messages to brands or marking them as spam to minimize future risks. Educational initiatives like cyber awareness training and staying updated on the latest cybersecurity scams were seen as vital tools for preventing smishing. As one brand stated, 
\begin{displayquote}
    “\textit{Complete our Cyber Fraud \& Secure Online Banking training in the Support \& Community Section of J.P. Morgan Access (J.P. Morgan)"}
\end{displayquote}

Finally, brands advocated for vigilance and skepticism when encountering unexpected requests, such as demands for payment for hospital or utility bills. Practices like constant account activity monitoring, recognizing red flags, and trusting instincts were also highlighted as effective deterrents. One brand succinctly captured this sentiment, stating, 
\begin{displayquote}
    “\textit{Regularly check your account activity for any suspicious transactions and contact us immediately about any suspicious or erroneous wires. (J.P. Morgan)"}
\end{displayquote}
 and another brand mentioned, 
 \begin{displayquote}
      “\textit{Be skeptical of unexpected calls, emails, or messages asking for sensitive information. Trust your instincts and take precautions to ensure you’re protected. You can stay one step ahead of scammers by always verifying what you’re being told—even if it means hanging up or ignoring a message until you’re sure. (Navy Federal)"}
 \end{displayquote}

Out of the 149 brands that we analyzed, only 69\% of the brands provided advice on how to prevent smishing. The brands that provided the content on how to prevent smishing had guidelines related to the above themes. The remaining 31\% of the brands had no mention of steps to identify smishing. The customers of these brands unfortunately are not educated about the scenario, even though these were popular brands whose names were part of previously reported smishing scams. This lack of content supports that lack of awareness increases the chances of a victim falling into a smish trap as the victim is not educated enough on smishing prevention.

\textbf{RQ3 Summary:} Even though different brands provide different advice on how to report smishing and steps after becoming a victim, they have a lot in common. The pieces of advice range from trusting user instincts, and blocking the smishing account to reporting smishing to law enforcement agencies. All the brands who provided the advice related to smishing reporting recommend either reporting or taking some level of action in one way or another.
\section{Discussion}

\subsection{General Discussion}
In this study, we analyzed the smishing awareness and prevention education provided by 149 distinct brands. We leveraged encoding techniques and identified relevant themes that most of the brands highlighted. We discovered that a greater percentage of the brands don't mention relevant or sufficient information on smishing, even though there have been smishing attack attempts using the brand's name. Below, we discuss and conclude the findings from each of the categories that we analyzed qualitatively: \\

\textbf{Smishing Definitions:} Brands differed in how they defined smishing to their customers. They most commonly portrayed smishing as an act of deception involving text messages, a malicious link, and a feeling of fear or urgency. There were also references to technical definitions like a type of social engineering attack, SMS phishing, and spam text. Some of the brands did not even mention the definition for smishing. This inconsistency highlights the need for a standardized, user-friendly definition. Brands should consider adopting definitions from authoritative sources such as CISA, which offer their own definition on smishing~\cite{cisaDefinition}. Using federal agency definitions as a foundation promotes consistency across platforms, while still allowing brands to tailor their messaging to the specific threats their users may face.

\textbf{Prevention Advice: } The prevention advice analyzed from different brands implied a mix of user-focused strategies. For example, always being cautious while interacting with an SMS and always verifying the sender and any links before reacting to an SMS. On the contrary, technology-focused strategies like two-factor authentication, constantly tracking account activity, and using firewalls were not emphasized much. While user behavior plays an integral part in a victim falling into a smishing attack~\cite{blancaflor2021phishing}, technical safeguards are also important to increase resistance against falling for a smish.

\textbf{Reporting Fraud Instructions: } Different brands had different ways of dealing with smishing attack reports. While some brands suggested directly reporting a smishing attempt to the respective brand, others suggested seeking government aid and reporting to anti-phishing organizations or local consumer protection agencies. We observed that not many brands (n=6) mentioned reporting the smish message to 7726, which is an official government helpline for reporting smishing messages. Some brands even mentioned reporting the message to the respective provider. This highlights the need for brands to mention both government resources and the brand's resources (if there are any) for reporting a smish message, to ensure a better and more detailed scrutiny of the smish attempt.

\textbf{Identifying Smishing:} Different indicators were suggested by different brands to identify a smishing attempt. The most common themes included SMS requesting personal information and SMS portraying a sense of urgency, both of which prompt the user to take quick action. Unexpected links from senders, cross-checking with the brand, and trusting personal instincts were also common themes. Additional markers highlighted were grammatical errors, fake offers, and money lending. All these prominent themes indicate that potential victims should always be alert and on their toes. Having cyber and smishing education also plays a significant role in identifying a potential smish message. 

\textbf{Steps to take after becoming a victim:} While there were not many brands that included the steps to take after becoming a victim, some did suggest detailed steps. Informing the service provider, reporting to law enforcement, and changing login credentials were the most prominent themes in the qualitative analysis. A few brands also suggested monitoring the account and installing antivirus software to neutralize the attack and prevent any potential loss. Less common themes included blocking scam numbers, remaining calm, and seeking advice from the respective brand. While all of these suggestions are valid, there is a need to define the steps to take after becoming a victim in a more streamlined manner. The steps should cover protecting accounts and given information, appropriately reporting the smishing attack, and measures to take to avoid a potential attack in the future.

\subsection{Recommendations}
After conducting a comprehensive qualitative analysis of smishing awareness advice provided by major tech companies and financial institutions, several recommendations emerge to enhance cybersecurity resilience and protect individuals and organizations from smishing attacks:

\textbf{Standardization of smishing Definitions:} 
Brands primarily defined smishing as a “deceptive message" or “a message impersonating a brand or a person." However, “SMS containing a short URL" or “social engineering attack" wasn't mentioned much. This leads to ambiguity and incomplete education for the brands' customers. There is a need for greater standardization in cybersecurity and smishing education provided across different brands. While variations tailored to each brand’s content are noted, aligning definitions with industry best practices can promote consistency and clarity in anti-smishing efforts. Further, expanding the scope to include critical identifiers like “short URL" would also provide comprehensive exposure to smishing education and common smishing tactics for potential victims.

\textbf{Simplifying Reporting Methods:} Reporting methods like “Report to respective brand" and “Contact FTC" were widely suggested. However, using official channels or reporting to 7726 was not mentioned much. This causes a problem in streamlining a reporting mechanism for the victim, who has to navigate different and complex reporting instructions depending on the brand. This fragmentation aligns with recent calls for improving fraud reporting infrastructure, as Button et al.~\cite{button2025policing} identified inadequate reporting and intelligence coordination as critical barriers to effective fraud response, particularly for cross-border threats. We suggest that brands should include a more streamlined reporting method and incorporate government and carrier reporting methods for a more efficient reporting process. Another possible solution could be, popular phone companies (like Apple and Google) or network providers (like AT\&T and Verizon) could inform the respective brand whenever their users report a smishing attempt to them.

\textbf{Improve Smishing Identification and Prevention Education:} Identification methods such as “SMS creates a sense of urgency" and “SMS contains unexpected links" were widely mentioned across the websites of different brands. However, important details like “shortened URL" or “includes attachments" were not covered. For prevention advice, basic yet vital instructions such as “Don't click any links" and “Always be cautious" were briefly covered across most brand websites. However, we observed that tips like “Check website URL thoroughly" and “Trust instincts" received limited mention.

Based on these findings, we recommend that brands provide more streamlined yet detailed education and content on smishing identification and prevention methods. Prior research also shows that more actionable and practical approach of providing cyber education is recommended~\cite{bada2019cybersecurityawarenesscampaigns}. Brands can utilize modern multimedia tools to design interactive educational content for better user understanding. To imply this, even if a game-based approach works better as suggested by Khan et al., brands should invest the resources to create this particular awareness~\cite{khan2022game}. Streamlining these aspects would help potential victims be cautious and detect smishing in its early stages. This would further assist victims in reporting the smish attempt and securing their personal information.

\textbf{Provide examples and video tutorials:} Only 35\% of the 149 brands that we explored provided smishing examples, and less than 1\% provided a video tutorial related to smishing. Examples and visual content could help a potential victim in numerous ways, such as: What does a smish message look like? What does a malicious link look like? What red flags should be looked for when inspecting a text message? How to report a smish message? How is malware installed through a link?

Reading plain text content can be boring for users and presents a greater challenge in maintaining their attention~\cite{abawajy2014user}. Instead, we recommend that brands invest in educational videos and interactive content for their users to better understand this type of scam. Interactive content like quizzes and puzzles could greatly attract more users to learn about smishing. Examples of previous smishing attempts would also help users understand scenario-based smishing. For example, in a bank scam, the scenarios would include asking for OTP or bank passwords. This scenario education would help to immediately alert the user when they see a text message with a similar scenario. It would also provide the user with a real-life walkthrough of a smish attempt.

Overall, by implementing these recommendations, brands can strengthen their smishing awareness efforts and effectively mitigate the risks posed by this prevalent form of cybercrime. Brands can increase their defenses against smishing attacks and empower users to recognize and respond to potential risks more effectively.

\subsection{Limitations}
Despite our best efforts, this research does have areas for improvement and elaboration. For example, while we feel our collection of meaningful data is solid, there are ways to improve it to reflect further findings. Also, the number of brands in each category is not equally divided. For example, we analyzed 25 Technology \& Software brands but only 4 Entertainment/Streaming brands. This also made it difficult to analyze and compare data across different categories. We selected brands that have already been found to be targeted by phishing and smishing attacks from Phishtank and Smishtank. To our surprise, we found that some of these brands did not have a webpage to warn customers about phishing or smishing, which is most concerning. The research team decided to take content from web pages from brands listed on Phishtank and Smishtank. There are possible chances we could have missed a few brands that are popular in smishing scams.

Additionally, we are analyzing content from web pages whose companies are based in the United States. There is an opportunity for other works to analyze corporations on a global scale or a more isolated approach using a different country as the company’s base of operations. Companies are often held to different standards depending on which country they operate under, both publicly and legally. Therefore, it is reasonable to inquire whether this affects preventive advice or lack thereof.

Another limitation is the use of publicly available content. These companies may engage in more activities to prevent customers from becoming victims of smishing; however, the research team decided to exclusively analyze content from reputable web pages as it can be difficult to find other such programs by the company.This means the study is limited to the brand awareness reflected on their websites, while companies may have employed other strategies, such as videos, awareness campaigns, or social media outreach to educate customers. Another thing one must acknowledge is that this research is reliant on this public information and does not have access to any private preventative advice that may be taking place. This may manifest itself in the form of internal company communication or training programs, which could be part of a larger effort of combating SMS phishing.

During the encoding process, we came up with the theme ‘explicit definition,’ and we determined that mentioning that phishing can be done through SMS constituted or was enough to be considered an explicit definition. However, upon closer inspection, it can be argued that this is also a missed opportunity for further analysis. It can be valuable to differentiate and analyze how frequently these web pages actually mention smishing by name rather than passively mentioning it as an avenue through which phishing is done. If we were to repeat this study, this change would certainly be employed.

\section{Conclusion}
Smishing attacks are rising over time. Brands are taking initiatives to provide awareness advice to their customers to reduce these attacks. In our study, we analyzed advice from 149 various brands that are most likely to be targeted by smishers and revealed several themes or strategies. We found that less than 50\% of brands mentioned smishing on their website and less than 1\% had a video tutorial. Smishing attacks are increasing exponentially day by day, and the lack of smishing awareness is one of the major reasons victims fall for a smish attack. Thus, we recommend that brands educate their customers in an interactive and streamlined manner about smishing so there is enough awareness among potential victims to combat it.

\label{sec:conclusion}

\bibliographystyle{plain}
\bibliography{main}

\section*{Appendix}

\section{Qualitative Codes}
\begin{itemize}
    \item Prevention Advice (46) \\
    \textit{Don't click any links or open any attachments (85), 
    Always be cautious when interacting with SMS (81), 
    Always check and verify the legitimacy of the sender's number, sms text and link included (80), 
    Don't share any personal information (73), 
    Do not respond to unverified senders or act quickly (36), 
    Stay thoroughly informed about the latest cybersecurity scams (32), 
    Only send money/pay after verifying the sender authenticity at least twice (28), 
    Contact the company directly to verify any suspicious text (26), 
    Don't share account related information (25), 
    Install latest version of antivirus software (24), 
    Avoid phone calls from unknown or suspicious callers (23), 
    Always double check gift card transactions (21), 
    Only use official website/app to perform or verify any action (20), 
    Maintain two-factor authentication and use strong passwords (17), 
    Report any suspicious message to the respective brand (17), 
    Check website URL and information thoroughly before entering any information (14), 
    Check grammar of the SMS (12), 
    Block and delete suspicious texts for avoiding future smishing attempts (8),
    Set alerts (8), 
    Sender Verification (7), 
    Suspicious Messages (7), 
    Cyber Awareness Training (6), 
    Avoid Text Messages (6), 
    Mark suspicious messages as spam (5), 
    Vigilance (4), 
    Unexpected debt collection (3), 
    Unexpected Hospital bills (3), 
    Skepticism (3), 
    Utility bills (3), 
    Avoid Installing Programs (2), 
    Constantly track your account activity (2), 
    Company sends SMS from only Specific Phone Numbers (1), 
    Fake Deals (1), 
    Firewalls (1), 
    Ignore and Delete (1), 
    Information Gathering (1), 
    Log-in credentials protection (1), 
    Lookout for impersonal greetings (1), 
    Not asking to delete app (1), 
    Enroll to Paperless Statements (1), 
    Public Computer Avoidance (1), 
    Red Flags (1), 
    Seek Advice (1), 
    Sensitive Information (1), 
    Session Security Restrictions (1), 
    Trust Instincts (1), 
    }

    \item Smishing Definitions (20) \\
    \textit{Sending deceptive message(23), 
    SMS requesting personal information (18), 
    Impersonates a brand or person (14), 
    Sending malicious links over SMS (12), 
    SMS Phishing (10), 
    Electronic fraud activity (9), 
    SMS creating urgency (9), 
    Phishing (8), 
    Sending fake information over SMS (8), 
    A type of cyberattack (6), 
    Asking to take immediate action (3), 
    Asks to install malware (3), 
    Type of social engineering attack (3), 
    Manipulate (2), 
    Scammers (2), 
    Threats (2), 
    Trickery (2), 
    Trustworthy entity (2), 
    Short URL (1), 
    Text Phishing (1)
    }
    
    \item Report Fraud Instructions (24) \\
    \textit{Report the SMS to the respective brand (68), 
    Contact FTC (14), 
    Contact government authorities (13), 
    Forward to Anti-Phishing Organizations (13), 
    Consumer Protection Agencies (8), 
    Online Reporting Platforms (7), 
    Report to cyber crime division (6), 
    Report to provider (6), 
    Forward SMS to 7726 (5), 
    FBI (4), 
    Internet Crime Complaint Center (4), 
    Report as Spam (3), 
    Share Screenshot (3), 
    IRS (2), 
    Local Law Enforcement (2), 
    Specific Information Required (2), 
    Consumer Sentinel Network database (1), 
    Delete Message (1), 
    Include all scam details (1), 
    Official Channels (1), 
    Platform Support (1), 
    Preventing Spread (1), 
    Screenshot Submission (1), 
    Timestamp (1)
    }
    
    \item How to Identify Smishing (24) \\
      \textit{ Ask for personal information(36), A sense of urgency(26), Unexpected links from sender(24), Trust your instincts(21), Misspellings or grammatical errors(17), Fake offers or promotions(15), Mimic trustworthy brands(13), Ask for funds(10), Cross checking with the brand(10), Include attachments (8), Unknown sender information (7), Impose threats(6), Mismatched links (6), Include third party(5), Fake delivery notifications(3), Ask to download an app or software (2), Provides phone number (2), Sender information hidden (2), Caution exercise (1), Directs to another website (1), Shortened URL(1),Emotional Appeal (1), Subtle differences with original brand(1), Variation of email domains(1),   }
    
    \item Steps to Take After Becoming Victim (21) \\
    \textit{ Inform and forward the relevant details to service provider(39), Report Law Enforcement (25), Change password and log out from devices (23),Account monitoring(22), Use anti-virus in the affected device (15), Do not engage(13), Check for identity theft (10), Inform Fraud Detection(10), Report to mobile carrier(10), Exercise caution for any future phishing (9), Educate yourself and other about cybersecurity(9), Take quick actions (8), Stop interaction if already engaged(8), Enable Two factor Verification(7), Inform friends and family(4), Learn from the provider website on how to secure account(4), Update email security(2), Provide written request to provider if necessary(2),Be calm(1), Block scam numbers (1), Research about phishing(1)  }
    
\end{itemize}

\section{Brands}
\begin{itemize}
    \item Technology \& Software (25):
    \textit{Adobe, Dropbox, ibm, accenture , Tech Mahindra, Cognizant , Intel, Accurint, Apple, Outlook, Microsoft, Google, Dynata, Verasight, Glide, HP, Eventbrite, Qualtrics, Samsung, Indeed, Intuit, Lenovo, Netsuite, Rackspace, Salesforce}
    \item Retail \& E-commerce (23):
    \textit{Target, Shopify, Ebay, Craigslist, Costco, BestBuy, Kohl's, JCPenny, Macy's, Amazon , walmart, walmart, American Greetings, Home Depot, Lowe's, Sam's Club, Best Buy, TJ Maxx, Kohl's, Gap Inc, Temu, Reebok, Rakuten}
    \item Finance \& Banking (23):
    \textit{Visa, American Express, CitiBank, Wells Fargo, Capital One, Discover, USBank, Synchrony, Navy Federal, J.P. Morgan, Bank of America, Chase Bank, ABSA Bank, Alliance Bank, Bank of the West, Barclays, Regions Bank, Navy Federal Credit Union, Deutsche Bank, US Bank, PNC Bank, Santander, Robinhood}
    \item Telecommunications (7):
    \textit{MetroPCS, Xfinity, Consumer Cellular, Viasaat, Verizon, AT\&T, Xfinity}
    \item Social Media (6):
    \textit{Instagram, TikTok, Reddit, Whatsapp, X, Snapchat}
    \item Automotive \& Ground Travel (6):
    \textit{Toyota, Uber, Lyft, Ford, Tesla, FasTrak}
    \item Cybersecurity \& VPN (6):
    \textit{NordVPN, Malwarebytes, Hut Six, Cofense, Palo Alto Network, Duo Security}
    \item Food \& Beverages (5):
    \textit{Coca Cola , Pepsi, Burger King, McDonald's, Domino's}
    \item Gaming (5):
    \textit{Arenanet, Blizzard Entertainment, GameStop, Steam, Zynga}
    \item Payment Services (5):
    \textit{Paypal, Stripe, Venmo, Square, Western Union}
    \item Logistics \& Delivery (4):
    \textit{UPS, FedEx, DHL, USPS}
    \item Health \& Pharmaceuticals (4):
    \textit{CVS, Walgreens, Aetna, Walgreens}
    \item Insurance (4):
    \textit{Allstate, StateFarm, USAA, Geico}
    \item Housing \& Utilities (4):
    \textit{SDGE, Shell, ExxonMobil, Chevron}
    \item Entertainment/Streaming (4):
    \textit{Netflix, Youtube, Disney, Hulu}
    \item Dating Platform (3):
    \textit{Coffee Meets Bagel , Bumble, Hinge}
    \item Air Travel (3):
    \textit{Delta Airlines, Virgin Airlines, American Airlines}
    \item Government (3):
    \textit{IRS, Electronic Benefit Transfer, Employee Development Department}
    \item News \& Information (2):
    \textit{Yahoo, WHO}
    \item Finance \& Investments (2):
    \textit{Berkshire Hathaway, Blackrock}
    \item Cryptocurrency (2):
    \textit{CoinBase, Binance} 
    \item Education (1):
    \textit{CSU Northridge}
    \item Non-Profit \& Advocacy (1):
    \textit{Vote.org}
    \item Food Delivery (1):
    \textit{GrubHub}

\end{itemize}

\end{document}